\documentclass[a4paper,12pt]{article}

\usepackage{amsmath}
\usepackage{amssymb}
\usepackage{latexsym}

\begin{document}

\newcommand{\im}{\mathrm{i}}

\renewcommand{\title}[1]{\null\vspace{10mm}\noindent{\Large{\bf #1}}\vspace{5mm}}
\newcommand{\authors}[1]{\noindent{\large #1}\vspace{15mm}}
\newcommand{\address}[1]{{\center{\noindent #1\vspace{0mm}}}}
\renewcommand{\abstract}[1]{\vspace{17mm}
\noindent{\small{\em Abstract.} #1}\vspace{2mm}}

\begin{titlepage}
\begin{center}
\hspace*{\fill}{{\normalsize \begin{tabular}{l}
                              {\sf hep-th/0202093}\\
                              {\sf REF. TUW 02-04}\\
                              {\sf REF. UWThPh-2002-07}
                             \end{tabular}   }}

\title{\vspace{2mm} IR-singularities in Noncommutative Perturbative Dynamics}

\vspace{10mm}

\authors {  \Large{J. M. Grimstrup$^{1}$, H. Grosse$^{2}$, L. Popp$^{3}$, V. Putz$^{4}$, M.~Schweda$^{5}$, M. Wickenhauser$^{6}$, R. Wulkenhaar$^{7}$ }}    \vspace{-20mm}

\vspace{10mm}
       
\address{$^{1,3,5,6}$  Institut f\"ur Theoretische Physik, Technische Universit\"at Wien\\
      Wiedner Hauptstra\ss e 8--10, A-1040 Wien, Austria}
\address{$^{2}$  Institut f\"ur Theoretische Physik, Universit\"at Wien\\Boltzmanngasse 5, A-1090 Wien, Austria   }
\address{$^{4,7}$} Max-Planck-Institute for Mathematics in the Sciences \\
 Inselstra\ss e 22--26, 04103 Leipzig, Germany

\footnotetext[1]{Work supported by The Danish Research Agency.}

\footnotetext[3]{Work supported in part by ``Fonds zur F\"orderung der Wissenschaftlichen Forschung'' (FWF) under contract P13125-PHY.}

\footnotetext[4]{Work supported by ``Fonds zur F\"orderung der Wissenschaftlichen Forschung'' (FWF) under contract P15015-TPH.}

\footnotetext[7]{Schloe\ss mann Fellow.}
       
\end{center} 
\thispagestyle{empty}
\begin{center}
\begin{minipage}{12cm}

\vspace{10mm}

{\it Abstract.}   We analyse the IR-singularities that appear in a noncommutative scalar quantum field theory. We demonstrate with the help of the effective action and an appropriate field redefinition that no IR-singularities appear in the quadratic part at one-loop order. No new degrees of freedom are needed to describe the UV/IR-mixing.

\end{minipage}\end{center}
\end{titlepage}

\section{Introduction}
In this letter we discuss an alternative approach to the problem of the so-called UV/IR-mixing in noncommutative Euclidean scalar $\phi^4$-theory. The classical  action is given by 
\begin{equation}
  \label{action}
  S = \int d^4\!x \, \left( \frac{1}{2} (\partial_\mu \phi \partial^\mu \phi 
    + m^2 \phi^2)+
      \frac{g^2}{4!} \phi \star \phi \star \phi \star \phi \right),
\end{equation}
with the noncommutative, associative star product
\begin{gather}
  (f \star g)(x) := \int \frac{d^4\!k}{(2 \pi)^4} \int d^4\!y \, 
    f(x + \frac{1}{2} \theta \cdot k) \, g(x + y) \mathrm{e}^{\im k \cdot y}, 
      \nonumber \\
\intertext{where}
  (\theta \cdot k)^\mu = \theta^{\mu\nu} k_\nu, \quad k \cdot y = k^\mu y_\mu, 
\end{gather}
and $\theta_{\mu\nu}$ the constant, antisymmetric noncommutativity parameter.

The perturbative properties of such a noncommutative field model are studied in great detail in \cite{Minwalla:1999px}, \cite{VanRaamsdonk:2000rr}. The $\star$-product in the interaction term leads to a momentum-dependent phase factor associated with each vertex in a given Feynman diagram. These phases create two sorts of graphs: planar and nonplanar diagrams, originally proposed in \cite{Filk:dm}, where the usual UV-renormalization procedure is applied to the planar graphs. The nonplanar diagrams contain phase factors dependent on the internal momentum, that are associated with each crossing of lines in the graph. The rapid oscillations of these phases regulate the integrals and thus suppress any divergence, i.~e.\ an otherwise divergent graph becomes finite with an effective cutoff (at one-loop)
\begin{equation}
  \Lambda_\mathrm{eff} = \frac{1}{\sqrt{\theta_{\mu\nu} p^\nu \theta^{\mu\rho}
    p_\rho}} \equiv \frac{1}{\sqrt{\tilde{p}^2}}.
\end{equation}
Therefore, the original UV-divergence is replaced by an IR-singularity in the limit of vanishing $p$ implying $\Lambda_{\mathrm{eff}} \rightarrow \infty$ (for a rigorous analysis see \cite{Chepelev:2000hm}).

This presence of IR-singularities in massive theories suggests the need of new light degrees of freedom \cite{Minwalla:1999px}, \cite{VanRaamsdonk:2000rr}, whereby one can reproduce the quadratic IR-singularity in the two-point vertex function.

The aim of our short contribution is to present an alternative way of discussing these IR-singularities. In \cite{Bichl:2001cq}, \cite{redef} we have shown that the field redefinition originally proposed in \cite{Bichl:2001nf}, \cite{Bastianelli:1990ey}, \cite{Alfaro:1992cs} is very useful for the perturbative description of noncommutative $U(1)$ gauge field models. Therefore, it was quite natural to use an appropriate field redefinition also in the present case of a scalar model to analyze the IR-structure of an effective two-point vertex function at $O(g^2)$. Similar results have been derived in \cite{Griguolo:2001ez} in the context of Wilsonian RG and hard noncommutative loop resummation.

This letter is organized as follows. Section \ref{effect} is devoted to the presentation of known results in this field. In section \ref{refo} we describe our solution to the IR-problem of the effective two-point 1PI-function. Finally, there is a short conclusion and outlook.

\section{One-loop effective action} \label{effect}
In this section we briefly review the well known results of \cite{Minwalla:1999px}, \cite{VanRaamsdonk:2000rr} showing that the one-loop nonplanar graphs in the scalar noncommutative field theory on four-dimensional Euclidean space are convergent at generic values of external momenta due to rapid oscillations of the phase factors $\mathrm{e}^{\im p \times k}$, where $p$ is an external momentum, $k$ the loop momentum, and $k \times p = k_\mu \theta^{\mu\nu} p_\nu$. 

Considering the two-point vertex function in momentum space one observes that at the lowest order it is just the inverse propagator
\begin{equation}
  \Gamma_0^{(2)} = p^2 + m^2.
\end{equation}
When radiative corrections are taken into account, two graphs are relevant \cite{Minwalla:1999px}, \cite{VanRaamsdonk:2000rr}: the planar and the nonplanar diagram.  

In order to calculate these two contributions in momentum space one has to evaluate the following two integrals 
\begin{align}
  \label{seibergint}
  \Gamma^{(2)}_{1 \, pl} &= \frac{\hbar g^2}{3(2\pi)^4} \int \frac{d^4\!k}
    {k^2 + m^2},  \nonumber \\
  \Gamma^{(2)}_{1 \, npl} &= \frac{\hbar g^2}{12(2\pi)^4} \int \frac{d^4\!k}
    {k^2 + m^2} ( \mathrm{e}^{\im k \times p} + \mathrm{e}^{-\im k \times p} ).
\end{align}
For $\theta=0$ the above integrals are identical up to a factor 2.

The planar graph leads to the mass renormalization 
\begin{equation}
  M^2 = m^2 + \delta m^2,
\end{equation}
where $M$ is the renormalized physical mass and $\delta m^2$ corresponds to a regularized planar graph and diverges quadratically if the cutoff tends to infinity. The mass counterterm $\delta m^2$ cancels just the divergence of the first integral in \eqref{seibergint}.

Using Schwinger parametrization 
\begin{equation}
  \frac{1}{k^2+m^2} = \int_0^{\infty} d\alpha \, 
    \mathrm{e}^{-\alpha (k^2+m^2)}
\end{equation}
one can easily calculate the resulting Gaussian integrals. Regularizing by multiplication with $\mathrm{e}^{\frac{-1}{\Lambda^2 \alpha}}$ one gets the following results:
\begin{align}
  \label{seires}
  \Gamma^{(2)}_{1 \, pl} &= \frac{\hbar g^2}{48 \pi^2} \left( \Lambda^2 - m^2 
    \ln \frac{\Lambda^2}{m^2} + O(1) \right) \nonumber \\
  \Gamma^{(2)}_{1 \, npl} &= \frac{\hbar g^2}{96 \pi^2} \left( \Lambda_
    {\mathrm{eff}}^2 - m^2 \ln \frac{\Lambda_{\mathrm{eff}}^2}{m^2} + O(1) 
      \right),
\end{align}
where
\begin{equation}
  \label{eq:lambdaeff}
  \Lambda_{\mathrm{eff}}^2 = \frac{1}{\tilde{p}^2 + \frac{1}{\Lambda^2}}
\end{equation}
In \eqref{eq:lambdaeff} we have introduced the notation $\tilde{p}^\mu = \theta^{\mu\nu} p_\nu$.

In the limit $\Lambda \rightarrow \infty$, the nonplanar one-loop graph remains finite, i.~e.\ the noncommutativity acts as a regularization scheme.

The one-loop 1PI quadratic effective action becomes therefore
\begin{multline}
  \label{gammaeff}
  \Gamma^{(2)}_{\mathrm{eff}} = \int \frac{d^4 \! p}{(2 \pi)^4} \, \frac{1}{2}
    \phi(p) \phi(-p) 
      \left(p^2 + M^2 + \frac{\hbar g^2}{96 \pi^2 (\tilde{p}^2 + \frac{1}
        {\Lambda^2})} \right. \\
    - \left. \frac{\hbar g^2}{96 \pi^2} M^2 \ln \left( \frac{1}
     {M^2(\tilde{p}^2 + \frac{1}{\Lambda^2})} \right) + O(g^4) \right). 
\end{multline}
The first line of \eqref{seires} implies 
\begin{equation}
  \delta m^2 =  \hbar \left( \frac{g^2 \Lambda^2}{48 \pi^2} - \frac{g^2 m^2}
    {48 \pi^2} \ln \frac{\Lambda^2}{m^2} \right).
\end{equation}

As is explained in \cite{Minwalla:1999px} the limit $\Lambda \rightarrow \infty$ does not commute with the low momentum limit $p \rightarrow 0$ (IR-region)---the so-called \emph{UV/IR-mixing} of noncommutative quantum field theories.

\section{Reformulation} \label{refo}
One tries to obtain a ``new'' effective one-loop two-point vertex function 
\begin{multline}
  {\Gamma'}^{(2)}_{\mathrm{eff}} = \int \frac{d^4 \! p}{(2 \pi)^4} \, \phi(p) 
    \phi(-p) \Bigg( p^2 
      + M^2  \\
    - \frac{\hbar g^2}{96 \pi^2} M^2 \ln \left( \frac{1}{M^2 
      (\tilde{p}^2 + \frac{1}{\Lambda^2})} \right) + O(g^4) \Bigg)
\end{multline}
(without the factor $\tfrac{1}{\tilde{p}^2 + \tfrac{1}{\Lambda^2}}$) as a result of a field redefinition
\begin{equation}
  \phi(p) \rightarrow \phi(p) + f(p, \theta, \Lambda) \phi(p).
\end{equation}
A simple calculation shows that a solution $f(p)$ is of the following form
\begin{equation}
  f(p) = -\frac{1}{2} \frac{\hbar g^2}{96 \pi^2} \frac{1}{(p^2 + m^2)} 
   \frac{1}{\tilde{p}^2 + \frac{1}{\Lambda^2}},
\end{equation}
whence the redefinition of the field $\phi(x)$ in position space
\begin{equation}
  \label{xredef}
  \phi(x) \rightarrow \phi(x) - \frac{1}{2} \frac{\hbar g^2}{96 \pi^2} 
    \frac{1}{(\Box - m^2)}\frac{1}{\tilde{\partial}^2} \phi(x).
\end{equation}
Therefore, also the action \eqref{action} must be changed accordingly
\begin{multline}
  \label{sredef}
  S' = \int d^4 \! x \, \left( \frac{1}{2} \left( \partial_\mu \phi 
    \partial^\mu \phi + m^2 \phi^2 + \frac{\hbar g^2}{96 \pi^2} \phi 
      \frac{1}{\tilde{\partial}^2} \phi \right)  \right.\\ 
    + \left. \frac{g^2}{4!} \phi \star \phi \star \phi \star \phi + O(g^4) 
      \right).
\end{multline}
It is now straightforward to compute an ``IR-regular'' quadratic effective action up to the given order in $g^2$ with this new action, yielding
\begin{multline}
  {\Gamma'}^{{(2)}}_{\mathrm{eff}} = \int \frac{d^4 \! p}{(2 \pi)^4} \, 
    \phi(p) \phi(-p) 
      \left(p^2 + M^2 - \frac{\hbar g^2}{96 \pi^2} \left( \frac{1}
        {\tilde{p}^2} 
          - \frac{1}{\tilde{p}^2 + \frac{1}{\Lambda^2}} \right) \right.  \\ 
    - \left. \frac{\hbar g^2}{96 \pi^2} M^2 \ln \left(\frac{1}{M^2 ( 
      \tilde{p}^2 + \frac{1}{\Lambda^2} )} \right) + O(g^4) \right).
\end{multline}
In the limit $\Lambda \rightarrow \infty$ one arrives at
\begin{equation}
  {\Gamma'}^{{(2)}}_{\mathrm{eff}} = \int d^4 \! p \, \phi(p) \phi(-p) 
    \left(p^2 + M^2 - \frac{g^2}{96 \pi^2} M^2 \ln \frac{1}{M^2 \tilde{p}^2} 
      + O(g^4) \right),
\end{equation}
which does not contain any non-integrable IR-singularities \cite{Grosse:2000yy}.

At this point one has to make the following comments:
\begin{itemize}
\item Because of the last term in \eqref{sredef} it is clear that the problem of UV/IR-mixing is not solved by this simple field redefinition, since the problems have only been transferred from the 2-point function to higher n-point functions. In fact, at higher orders in $g$ the field redefinition produces a term proportional to 
\begin{equation}
 \frac{\hbar g^4}{(\Box - m^2)} \frac{1}{\tilde{\partial}^2} \phi^4,
\end{equation}
which induces new IR-singularities.
\item The correction term in the field redefinition \eqref{xredef} is of order $g^2$. Thus, the bare propagator (the free-field case being defined by $g=0$) remains unchanged:
  \begin{equation}
    \Delta(p) = \frac{1}{p^2 + m^2}.
  \end{equation}
\item The field redefinition \eqref{xredef} is nonlocal and induces also a nonlocal term in the action \eqref{sredef}. Such nonlocal field redefinitions are known to arise in non-Abelian gauge field models quantized in the axial gauge, where the redefinition must be compatible with BRST-symmetry \cite{Boresch:1998} (and references therein).
\item In order to reproduce the UV/IR-mixing the authors of \cite{Minwalla:1999px}, \cite{VanRaamsdonk:2000rr} have interpreted the IR-singularities in the nonplanar one-loop diagrams as tree level exchange of new light degrees of freedom. In our approach there is no need of introducing these degrees of freedom.
\item Since the dangerous term $\frac{\hbar g^2}{96 \pi^2} \frac{1}{\tilde{p}^2 + \frac{1}{\Lambda^2}}$ in \eqref{gammaeff} does not depend on the mass (physical or bare mass) the massless case is also well defined implying a field redefinition
  \begin{equation}
    \phi(x) \rightarrow  \phi(x) - \frac{1}{2} \frac{\hbar g^2}{96 \pi^2} 
      \frac{1}{\Box} \frac{1}{\tilde{\partial}^2} \phi(x)
  \end{equation}
leading to the following quadratic effective one-loop action
  \begin{equation}
    {\Gamma'}^{{(2)}}_{\mathrm{eff}} = \int \frac{d^4 \! p}{(2 \pi)^4} \, 
      \phi(p) \phi(-p) \left(p^2 -  \frac{\hbar g^2}{96 \pi^2} \left( 
        \frac{1}{\tilde{p}^2} - \frac{1}{\tilde{p}^2 + \frac{1}{\Lambda^2}} 
          \right) + O(g^4) \right)  
  \end{equation}
for finite $\Lambda^2$.
\end{itemize}

\section{Conclusion and outlook}
In this letter we have demonstrated that the (quadratic) IR-singularities appearing in the 2-point function of noncommutative $\phi^4$-theory may be shifted to higher  n-point functions. One could speculate if this method, initiating an infinite chain of field redefinitions, could in fact be used to totally remove the IR-singularities.

In any case, the above results are certainly a strong motivation to study noncommutative gauge theories, also plagued by IR-problems \cite{prep}. There are some substantial hints that a nonlocal field redefinition of the gauge field might possibly be an appropriate recipe for curing the IR-singularities of noncommutative Yang-Mills theories.

\end{document}